\documentclass{optica-article}

\journal{oe}
\articletype{Research Article}
\usepackage{lineno}
\usepackage{float}

\begin{document}
\title{Radiometric sensitivity and resolution of synthetic tracking imaging for orbital debris monitoring}
\author{Hasan Bahcivan\authormark{1,*}, David J. Brady\authormark{2} and Gordon C. Hageman\authormark{2}}

\noindent \authormark{1}OpticalX, LLC., Tracy, CA, 95304, USA\\
\authormark{2}College of Optical Sciences, University of Arizona, Tucson, AZ, 85751, USA\\
%\authormark{3}The authors contributed equally to this work.\\
%\authormark{*}opex@optica.org}
\medskip

\section{Abstract}
  We consider sampling and detection strategies for solar illuminated space debris. We argue that the lowest detectable debris cross section may be reduced by 10-100x by analysis of stacks of image frames collected at high rates rather than single frame data. In particular, instead of a pixel as a spatial region, the analysis is based on a "phase-space-pixel" which corresponds to an angular velocity and space region and whose intensity is computed by a weighted stacking of spatial pixels corresponding to a test debris trajectory within a  wide camera field-of-view (FOV). To isolate debris signals from background, the exposure time is set to match the time it takes a debris to transit through the instantaneous field of view. Debris signatures are detected by multiscale X-ray processing of the data cube. Radiometric analysis of line integrals shows that sub-cm objects in Low Earth Orbit can be detected and assigned full orbital parameters by this approach.

\section{Introduction}

 The Low Earth orbit (LEO), defined as the altitude range between 400-1000 km,  contains millions of debris particles that pose a significant collision risk to existing spacecraft in operation. For sub-cm objects, the NASA Orbital Debris Program Office conducted  studies between the calendar years 2014-2017 using special staring radar modes at the MIT Haystack Observatory in Westford, Massachusetts, and the Goldstone Solar System Radar near Barstow, California Goldstone, each system being able to detect (but not track) 5mm and 3mm debris, respectively, at altitudes below 1000 km \cite{kennedy+etal-2020}. Per hour, on average 15 (60) debris passed through the extremely narrow beam  of 0.058$^{\circ}$ for Haystack (for 0.037$^{\circ}$ for Goldstone). The 3mm debris flux was measured at 0.03 counts per km$^2$. If one were to construct a vertical net around the Earth at Goldstone's latitude of 35$^{\circ}$ extending between 400-1000 km and wait for 90 min (typical LEO periods), one would collect approximately 1 million individual sub-cm pieces.  Meanwhile, debris generating events appear to happen more frequently. The number of active satellites has been rapidly increasing as the commercialization of space from LEO to GEO is becoming more attractive for communication and navigation systems.  Considering in-flight breakups such as the 2007 Fengyun-1C satellite test, 2009 collision of Iridium 33 and Kosmos 2251, 2021 Russia test of an anti-satellite weapon, 2024 breakups of the communication satellite Intelsat 33e and the military weather satellite DMSP-5D2 F14, and numerous other accidental satellite fragmenting explosions, debris accumulation will surely accelerate.  Ultimately, the concern is that the number of objects in the Low Earth Orbit beyond a certain threshold will trigger the "Kessler effect" \cite{kessler-1978}  that a small number of collisions will trigger an unintended exponentially growing  avalanche of fragments, making LEO to GEO unusable.  When such an avalanche begins, physically de-orbiting such large number of debris is not feasible. The only option is orbit maneuvering, and it requires knowing the orbits of each of the  debris pieces as small as micrometers, hours, or days in advance. 

Tracking of sub-cm debris is significantly beyond current capabilities, however. As of September 2022, CelesTrak, a non-profit organization,  maintains a catalog of 54000 space objects with sizes approximately 5 cm and larger.  Considering that 99.3\% of debris is below 1 cm \cite{lee+etal-2020}, the CelesTrak catalog is just the beginning of a lengthy one for true Space Situational Awareness (SSA). 

One needs a simultaneous scan and stare capability to rapidly look for all the objects in a wide field of view (FOV) while maintaining detection sensitivity. Unfortunately, conventional RF and optical systems trade in FOV for sensitivity. Increasing aperture improves sensitivity in staring mode, but reduces FOV and detection rate. Moreover, minimum three "angles-only" observations at different times are necessary for orbit determination. The likelihood of an object passing through multiple narrow beams for OD is even lower.  A phased-array narrow-beam radar, e.g., the Advanced Modular Incoherent Scatter Radar (AMISR) \cite{valentic+etal-2013} can achieve a wide FOV but the sensitivity at each beam position will decrease by the number of beams. However, this is not true with a 3D computationally post-steering receive array, which does increase the sensitivity while maintaining the same FOV. On the other hand, the concept of large receive arrays is logistically prohibitive, as the number of receivers scales inversely with the 4th power of the debris diameter. Alternatively, sensitivity can be increased using more radar power; however, this requires increasing the already MW+ radar power by multiple orders of magnitude, which is logistically prohibitive or would be a serious environmental concern. A similar argument applies to existing optical systems built for astronomy. Large-aperture deep-space telescopes are extremely sensitive to detect small debris for population statistics, but the rate of detection is very small. Moreover, the angular position and angular rate information from the images is not sufficient for OD which requires linkage of multiple detections, the chances of which are small. Finally, Satellite Laser Ranging (SLR) systems are built for precise OD, however, a coarse trajectory is needed for pointing.   Just like radars, transmitting high power lidar beams to artificially illuminate debris is problematic.

We discuss here a concept passive optical technique, Space-time Projection Optical Tomography (SPOT), that overcomes the above scan vs. stare problem with the following elements: (1) a massive parallel camera system with a large FOV,  (2) massive computational power using  Graphical Processing Units (GPUS) that enables synthetic stare capability while rapidly searching a large space of phase-space trajectories in 4-6 dimensions,  (3) precise analytical description of debris motion along geodesics allowing astrometric search space optimization, and (4) the Sun illuminates all the millions of debris at the same time with its powerful $\sim$1.36 $kW/m^2$ radiation.  SPOT can look everywhere within the wide FOV of a survey telescope and can zoom in on each and every particle in the FOV with the sensitivity of a deep-space telescope. The key concept is integrating pixel values corresponding to a track that is precisely defined by orbital parameters. For the limited case of straight tracks, the technique is equivalent to that of 2D Radon transform or to that of 3D Radon transform \cite{toft-1996} if a line on a plane is integrated, as in X-ray transform. For 2D transform, two angular parameters and two translational parameters define a line. Here, however, the lines are analytically defined by the orbital two-line element set (TLE), observatory position and observation time. The lines are curved differently; short ones could be analytically described by low order polynomials, while long ones like those of MEO or GEO orbits may require certain spiral forms to describe. In this regard, the transform is best described as a generalized Hough transform \cite{hough-1962}.
 Interestingly, Hough was trying to automate the task of detecting and plotting the tracks of sub-atomic particles in bubble chamber photographs\cite{hart-2009}. In a collisionless plasma confined by a magnetic field, an energetic electron track can be described by the Lorentz force precisely, just like gravity describes free-fall motion for time intervals short enough to ignore thermospheric drag or other non-gravitational forces. Such analytical precision enables high synthetic exposures by integrating pixels along long track projections on the FOV.

As shown by Shell \cite{shell-2010}, a single aperture imaging system can detect and track decimeter-scale LEO debris illuminated by sunlight during the twilight hours. Several proof-of-concept single-aperture implementations based on single-pixel thresholding demonstrated detections as small as 10 cm.\cite{wagner+clausen-2021,wagner+etal-2015, hasenohr+etal-2016}.  Typically, debris detection is done by looking for streaks in single frames. Various algorithms are tasked with identifying any streaks on the focal plane that are detectable above the noise. Preceding those algorithms is data preparation to clean known objects, such as stars, by looking up star catalogs. The tracks are used to calculate the angular coordinates and the angular velocity and are defined as "tracklets". Multiple tracklets are combined over multiple sightings to determine an initial orbit as a TLE \cite{lue+etal-2019}. For better accuracy, the initial TLE is passed to a more sensitive telescope as cued object or to Satellite Laser Ranging (SLR) stations for precision orbit determination.

A growing literature  aims to increase the effective exposure time by integrating pixels over multiple frames under certain assumptions on the trajectories, e.g., moving target indicator,\cite{kostishack+etal-1980} multiple hypothesis testing \cite{zingarelli+etal-2014}, direct track-before-detect methods \cite{davey+etal-2013}. 
The Hough transform has been used to detect space debris tracks for SSA \cite{svedlow+semenza-1993, privett+etal-2019, jiang+etal-2022} as well as to clean sky survey databases such as the SuperCOSMOS Sky Survey by detecting and removing such tracks due to various linear phenomena. \cite{storkey+etal-2004}.
Ultimately, the object trajectories are determined by gravity alone over time scales short enough to ignore thermospheric drag. One can form a space of such trajectories depending on the bounds of the orbital regions or on prior orbital information on the object \cite{murphy-2018} and then apply very selective matched filters to the image database to search for space debris.  \cite{murphy+etal-2017} 

Filtering with such extended filter banks and ultimately with exact TLE projections presents an advantage to LEO observations with a wide FOV camera by tightly constraining the orbital parameters. Six parameters are needed to describe a Keplerian orbit. It is generally assumed that the observable states of an optical measurement are angle and angle rates, e.g.,  ${\bf x} \in \mathbb{R}^4$, and no information can be reliably used to determine the other two states representing angular acceleration. \cite{milani+etal-2004}  This leads to the definition of admissible regions or a region of hypotheses consistent with a particular observation $\mathbb{R}^4$. \cite{murphy-2018}  While the assumption $\mathbb{R}^4$ is largely true for GEO because of the large radial distance to the object and the very low angular rates that do not change over the relatively narrow FOV, LEO is very different. LEO has very high angular rates that change significantly from near zenith to near horizon. Such angular rate variations can be captured by a wide FOV camera design providing the other two undetermined states. Thus, optical measurements that enable ${\bf x} \in \mathbb{R}^6$ including angular acceleration significantly constrain the search space allowing the discovery of LEO objects with full orbital parameters.

 SPOT is designed to search the $\mathbb{R}^6$ space for the uncued discovery of sub-cm and larger geocentric objects by simultaneously achieving high sensitivity and high search rates using computationally intensive synthetic tracking algorithms.  The algorithm is analogous to computationally zooming in on runway sequential flashing lights, also known as "rabbit lights" in aviation. Since the debris is so faint,  the idea is to use a very short exposure time per frame (to minimize background)  while maximizing the signal exposure time by stacking only the pixels that contain the signal. This methodology is similar to the algorithms  successfully used to detect faint objects in the asteroid belt. \cite{shao+etal-2014, zhai+etal-2014, zhai+etal-2020, heinze+etal-2015, heinze+etal-2019} except that the time scales for Earth-orbiting objects are much faster.  For LEO, SPOT can stare at each debris for up to 10 mins, although the transit time through each pixel is as short as 10 ms. This synthetic maximization of the exposure time is key to detect sub-cm debris, while the ability to search in the wide FOV is key to cataloging large number of debris per unit time.

Below, we  review Shell's radiometric analysis for single-pixel SNR and then extend the calculations to multi-pixel integrals over the entire FOV to maximize the incorporation of all information from the measured images. We computed the cumulative SNR for the sample LEO trajectories and for various debris sizes.  We then discuss computational aspects of debris search. Finally, a brief performance analysis of a worldwide network such observatories is made to understand how multiple data sets can be combined to provide improved and timely tracking capability. 

\section{Single-pixel SNR dependency on exposure time}
The photoelectron content of a pixel is due to (1) $e_s=e_{s0}\tau_t$: signal photons acquired during the object transit time $\tau_{t}$ through the pixel, (2) $e_b=e_{b0}\tau_e$: background photons acquired during the exposure time $\tau_{e}$, (3) $e_n$: detector noise per reading, and (4) unwanted photons coming from other distinct objects such as stars and man-made objects. For simplicity, let us ignore (4) and assume that the best efforts are made to mask the associated pixels. Here, the radiometric values $e_{s0}$ and $e_{b0}$ are detailed in the next section. Then 
\begin{equation}
e_{total} = e_{s0}\tau_t + e_{b0}\tau_e + e_n
\label{etotal}
\end{equation}
For a particular observation, geometry $\tau_t$ is already determined for each pixel.  Now, we look for the best $\tau_e$ to maximize the single-pixel SNR.  For $\tau_{e} > \tau_{t}$, it is given by
\begin{equation}
SNR_{\tau_{e} \ge \tau_{t}} = \frac{e_{s0}\tau_t}{\sqrt{ e_{b0}\tau_e + e_n^2            }              }
\label{snr1}
\end{equation}
and maximizes when $\tau_{e} = \tau_{t}$. This is easy to see because, beyond the object transit time, the pixel is no longer collecting signal photons while continuing to accumulate background photons. However,  for $\tau_{e} < \tau_{t}$, 
\begin{equation}
SNR_{\tau_{e} \le \tau_{t}} = \frac{e_{s0}\tau_e}{\sqrt{ e_{b0}\tau_e + e_n^2            }              }
\label{snr2}
\end{equation}
Here again, the SNR maximizes when $\tau_{e} = \tau_{t}$ because it is an increasing function of $\tau_{e}$.
Therefore, the best SNR for a pixel box is achieved when the exposure time is equal to the transit time. We discuss below how the transit time depends on the sensor size and how too-small a transit time is a problem in the presence of read noise. Generally, the telescope and camera parameters are chosen with the read noise setting a lower bound on the tolerable background noise. As can be seen from Eq. \ref{snr1},  if the background noise is already much smaller than the read noise, further shortening the exposure time does not significantly enhance the SNR and the resulting higher frame unnecessarily adds to the data acquisition and processing burden.  

The line that pierces through pixels of a data cube, i.e. stacks of frames with each frame $\tau_e$ thick, is the projection of an Earth-centered Keplerian orbit. 
Each pixel box of the data cube on the line captures various fractions of the signal photoelectrons. Now, the problem of finding debris may be formalized as follows: determine if a line of pixels corresponding to a particular orbit is occupied. As we discussed above, that line of pixels is represented by a phase-space-pixel (PSP), a point in a six-dimensional space. It is a generalization of the 'vector-pixel' as used in \cite{heinze+etal-2015} for constant angular speed objects.

\section{Radiometric equation for a phase-space-pixel}
The number of signal ($e_s$) and background ($e_b$) photoelectrons from a single detector pixel is given, respectively, by \cite{shell-2010} 
\begin{eqnarray}
e_s &=& QE \cdot \tau \cdot A \cdot \tau_{atm} \cdot E_{RSO} \cdot \tau_{t} ~~~~~~~[e^-] \\
e_b &=& QE \cdot \tau \cdot A \cdot L_{B} \cdot \frac{x^2}{f^2} \cdot \tau_{e} ~~~~~~~[e^-]
\end{eqnarray}
where $QE$ is quantum efficient, $\tau$ is optical transmittance, $A$ is aperture area, $\tau_{atm}$ is atmospheric transmittance, $x$ is pixel size, $\tau_{t}$ is object transit time over an IFOV, $\tau_{e}$ is the exposure time, $L_{B}$ is the background radiance, and $f$ is focal length.

% Clarification on p versus rho may be needed below

The visual magnitude of the resident space object (RSO) $E_{RSO}$ is given by \cite{shell-2010}
\begin{equation}
E_{RSO} = 5.6 \times 10^{10} \cdot 10^{-0.4m_{sun}}\cdot \frac{d^2}{R^2}[0.25 \rho_{\rm spec} + \rho_{\rm diff}p_{\rm diff}(\psi)]
\end{equation}
where $m_{sun}$ is the visual magnitude of the Sun (-26.7), $d$ is the diameter of the RSO and $R$ is the range, $\rho_{\rm spec}$ and $\rho_{\rm diff}$ are specular and diffuse components of the RSO albedo ($\rho=$0.175 is equally split), the latter being a function of the solar phase angle $\psi$ as follows:
\begin{equation}
p_{\rm diff}(\psi) = \frac{2}{3\pi}[\sin(\psi) + (\pi-\psi)\cdot \cos(\psi)]
\end{equation}

%================================================%

The atmospheric background $L_{B0}$ at zenith (photons per second per meter squared per steradian) is given by
\begin{equation}
L_{B0}[i,j,k] = 5.6 \times 10^{10} \cdot 10^{-0.4m_b} \cdot (\frac{180}{\pi})^2 \cdot 3600^2 
\end{equation}
and $m_b$ is the background in units of visual magnitudes per square arc seconds. That is, it is a measure of visual magnitude spread out over a square arcsecond of the sky. For a given $m_b$, a smaller IFOV will result in a smaller background noise per pixel. A truly dark sky has  $m_b=$21.8 mag arcsec$^{-2}$. The sky at its zenith becomes darker by $\approx 0.4$ mag arcsec$^{-2}$ during the first six hours after the end of twilight (sun 6 degrees below). \cite{walker-1988}. However, the time variation will be ignored here. Instead, we will use the zenith angle dependence. To calculate the background at $z$ degrees away from the zenith, we used
\begin{equation}
L_B = L_{B0}[0.4 + 0.6 \cdot (1-0.96\sin^2z)^{-1/2}]
\end{equation}
This formula by Garstang \cite{garstang-1989} is based on the analysis by Roach and Meinel [1955] \cite{roach+meinel-1955}, in which 40\% of the average faint star background and 60\% come from airglow emissions at a height of 130 km. We compared this dependence on $z$ with the analytical form of Shell \cite{shell-2010} and found agreement within a magnitude of less than 1.

For detector noise, we will assume that there is no dark current and use an RMS value of 1 for read noise $e_n$. Then, the SNR for a single pixel is given by 
\begin{equation}
SNR = \frac{e_s}{\sqrt{e_b + e_n^2}}
\end{equation}

\begin{figure}[H]
\centering\includegraphics[width=7cm]{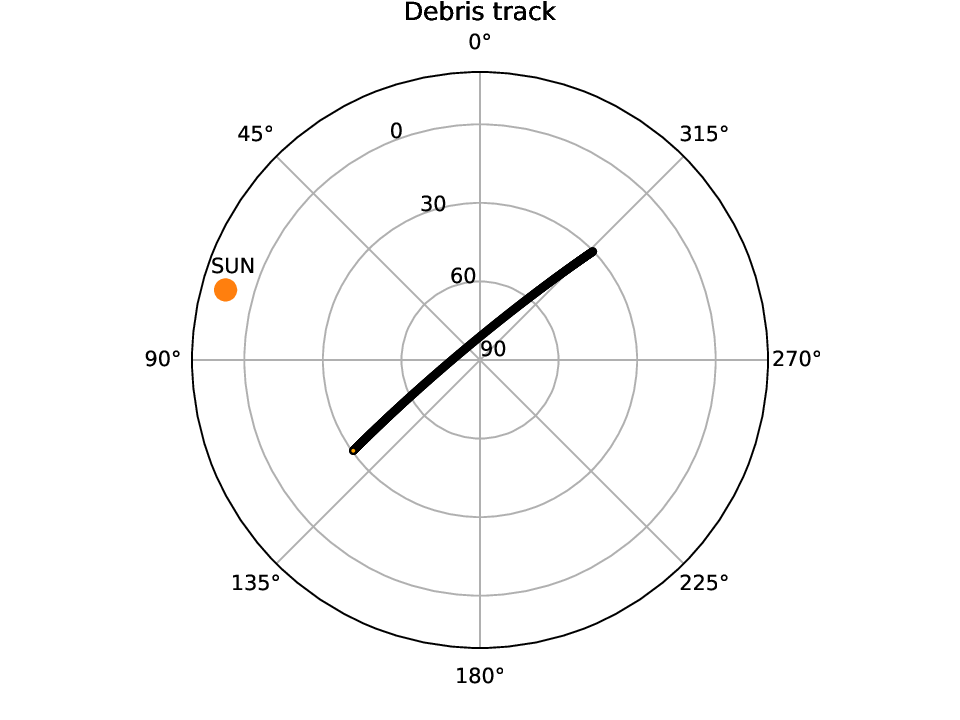}
\caption{TLE track = 10 cm}
\label{high_pass}
\end{figure}

Figure 1 shows a sample debris track corresponding to the following TLE: 
\begin{verbatim}
1 25544U 98067A 22241.38033696 .00006975 00000+0 13024-3 0  9998
2 25544 51.6450 333.9762 0003281 173.1914 302.2976 15.49962082356533
\end{verbatim}
The elevation and azimuth track is as seen from a location near San Francisco, CA, looking up. The $3$ min pass above $30^{\circ}$ elevation is during morning twilight; the Sun is at $\approx 11^{\circ}$  below the  horizon. This pass is chosen because of its high peak elevation, providing a simple geometry to analyze the effect of some of the radiometric parameters.

\begin{figure}[H]
\centering\includegraphics[width=7cm]{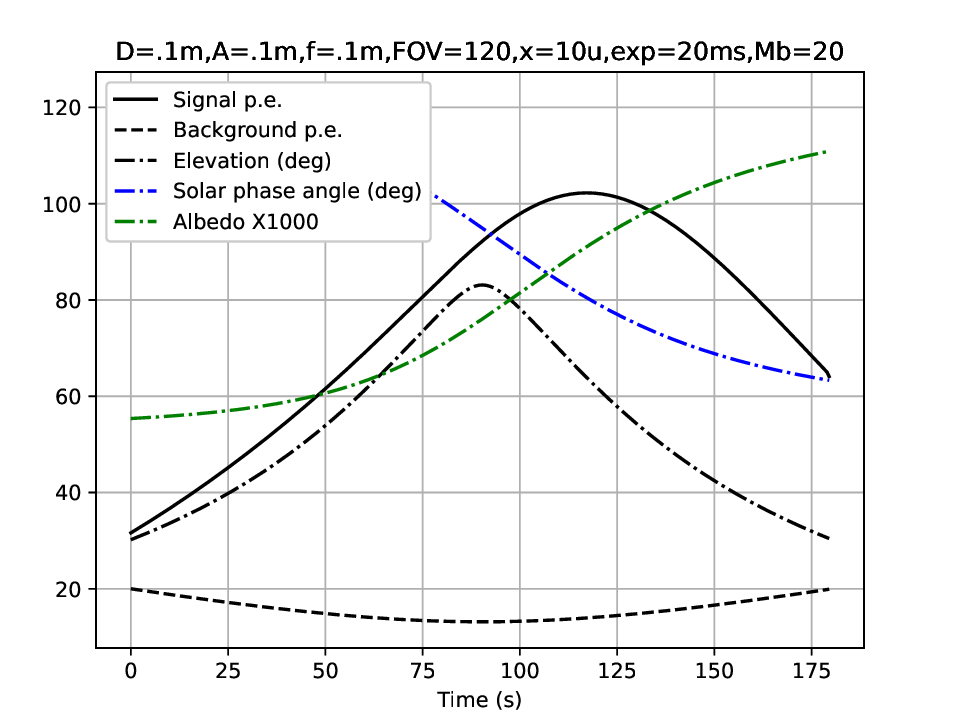}
\caption{Variation of single-pixel signal and background photoelectrons, debris elevation, solar phase angle, and albedo along the path shown in Figure 1. Here, the debris diameter $d=10 $ cm , aperture $A=10$ cm, focal length $f=10$ cm, and the pixel size $x=10~\mu m$. The exposure time for $e_b$ is $20$ ms. For $e_s$, pixel transit time is used to for effective exposure time.}
\end{figure}

Figure 2 shows the variation of the single-pixel signal and background electrons along the track for 10 cm debris. The exposure time $\tau_{e}$ is chosen as 20 ms for calculating the background noise while $\tau_t$ (for signal photons) is variable and given by the debris transit time through a pixel (dotted line). Note that the background is symmetric around the peak elevation (near 90 s) and significantly decreases with decreasing elevation as a result of equation 7. Thus, at low elevations and short exposure times, the read noise contribution  becomes dominant. Moreover, the skewness of the $e_s$ curve is due to the solar phase function given in Equation 9.

To obtain the radiometric equation for a PSP, we choose a hypothetical track $T[\theta_x(t), \phi_x(t)]$ as the angular trace of a hypothetical moving object viewed from a position $\bf x$. Next, we define a 3D pixel as $[i(t), j(t), k(t)]$ where $(i,j)$ are spatial indices  and $k$ is the image frame number, all step functions of $t$. We assume that there are no space or time gaps between pixels such that for each time $t$ the angular position of the object can be associated with a unique pixel. The task is to computationally zoom in on the moving object by integrating over the
pixels containing the object. 

% Clarification on '[cite]' and '[ref]' needed below
For simplicity, we assume that the camera is mechanically fixed, although it does not have to. We also assume that the camera orientation, pointing and lens distortion is calibrated to a small fraction of a pixel such that the effect of pointing uncertainties are negligible. This calibration can be achieved by matching the location of a large database of bright point sources such as the ESA’s Hipparcos space astrometry database, which provides the positions of more than one hundred thousand stars with high precision. Another source is the JPL DE421 ephemeris for the planets.

Note that a debris track $T$ can be accurately described by models including gravity and non-gravitational forces over the FOV such that its position perturbation due to unknown drag forces is much smaller than the physical dimensions of a 3D pixel. For example, the maximum deceleration due to the atmospheric drag of a 10 cm cube object with a density of 1$\rm g/cm^3$  (Area-to-Mass-ratio (AMR) is 0.01) at 780 km altitude is $10^{-9.5} {\rm km/s}^2$ \cite{montenbruck+gill-2000}. The position perturbation from the beginning to the end of a 1000 s long track will be 15 cm. These perturbations for a cm cube and a mm cube will be 1.5 m and 15 m, respectively, still small numbers compared to the 25-100 m IFOV footprint of the camera pixel at 780 km. Drag perturbations are modeled into orbit predictions; thus, the effect of unknown forces is minuscule and the effect on the accuracy of FOV projections is negligible.

%===================================================%

Continuing with the radiometric analysis, instead of SNR thresholding of a single-pixel for detection as is done in Shell\cite{shell-2010}, we SNR threshold the phase-space pixel, a weighted integral of the pixel values corresponding to a track.  Then, the line integral of the signal photoelectrons $e_s^L$ can be expressed as:
\begin{equation}
e_s^L = QE \cdot A \cdot \tau \int_a^b \tau_{atm}(t)E_{RSO}(t) dt
\end{equation}
where $\tau_{atm}(t)$ and 
$E_{RSO}(t)$ are now time-dependent because the range $R$, the solar phase angle $\Phi$, and the atmospheric penetration angle change as the debris moves. Note that no "pixellation" exists in the above integral; the assumption is that the selection of the pixel boxes $(i,j,k)$ will contain 100$\%$ of the signal photoelectrons.

The line integral of the background noise is given by 
\begin{equation}
e_b^L = QE \cdot A \cdot \tau \cdot \frac{x^2}{f^2}\sum_{(i,j,k)\in T}L_B[i,j,k] \tau_e[i,j,k]
\end{equation}
where $\tau_e[i,j,k]$ is the exposure time of the pixel $[i,j,k]$.
Although pixel-dependent exposure time is not expected from a single camera, for multiple cameras covering low- and high-altitude, variable exposure time may be used.
Note that, in contrast to $e_s^L$ which is a continuous integral, $e_b^L$ is a discrete integral. The main reason for this is that $\tau_t$ is a continuously changing variable while $\tau_e$ is preset as a system parameter. 

Similarly, the line integral of the read noise is a discrete sum,
\begin{equation}
e_n^L = \sqrt{\sum_{(i,j,k) \in T}e_n[i,j,k]}
\end{equation}
where we made $e_n$ a function of the location of the pixels in case multiple cameras with different read noise characteristics are used.

Then, the line integral SNR is given by
\begin{equation}
    SNR^{L} = \frac{e_s^L}{\sqrt{e_b^L+(e_n^L)^2}}
\end{equation}

\begin{figure}[H]
\centering\includegraphics[width=7cm]{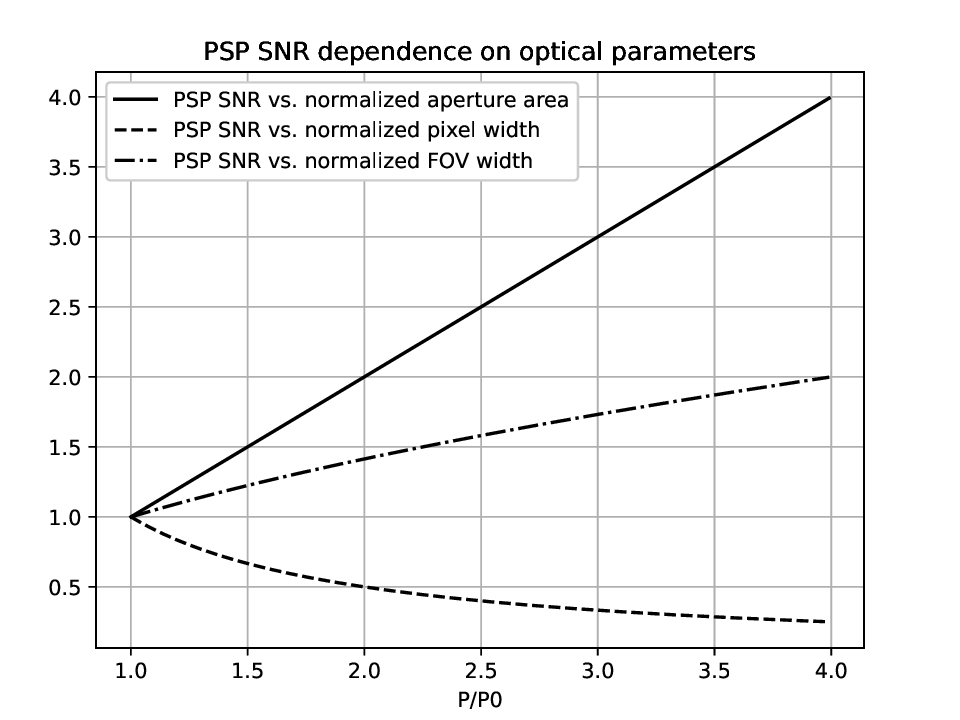}
\caption{PSP SNR dependence on the normalized system parameters: aperture area, pixel size and FOV extent. The focal length is set equal to the aperture diameter. When changing the aperture area,  FOV is kept constant by changing the number of telescopes. Read noise is ignored.}
\label{snr_dependence}
\end{figure}
Figure \ref{snr_dependence} shows how the PSP SNR changes as a function of normalized system parameters. If the read noise is negligible compared to the background noise, we arrive at the following analysis. Assume that the system captures a fixed duration of a debris track, which means a fixed FOV.  If the aperture diameter and focal length increase by 2, $SNR^L$ increases by 4. Note that $e_s^L$ increases by 4, but $e_b^L$ remains the same. This is so because although the number of pixels along the track increases by two due to the halving of IFOV, the exposure time $\tau_e$ will be halved to match the halved transit time through a pixel. Then, since $e_b^L$ is proportional to $A$ and IFOV$^2$, it will remain the same. 

If the read noise is the dominant source of noise, we arrive at a different analysis. If the aperture diameter and focal length increase by 2, $SNR^L$ increases by $2\sqrt{2}$. Note that $e_s^L$ increases by 4 but $(e_n^L)^2$ increases by 2, because twice the number of pixels is read.

\begin{figure}[H]
\centering\includegraphics[width=7cm]{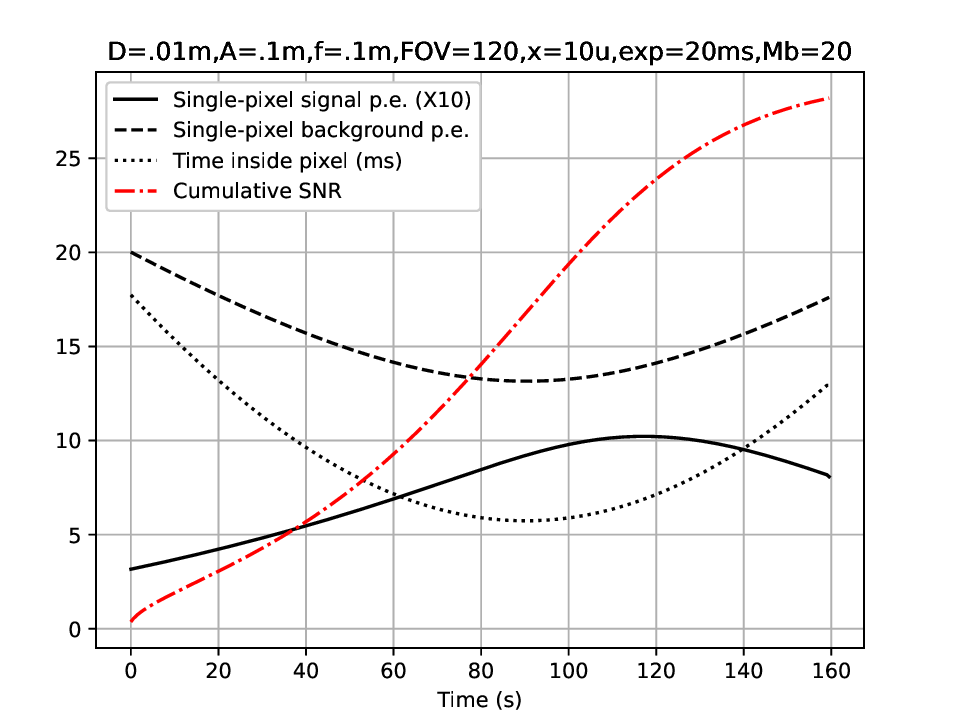}
\caption{Cumulative $SNR^L$ (red dashed line) for the debris pass shown in Figure 1 but for a 1 cm debris. Also shown are single-pixel SNR (solid black line), background (dashed black line) and time spent inside a pixel (dotted line). The optical parameters are: $d_{\rm deb}=1$ cm, $d_{\rm aperture}=10$ cm,$f=10$ cm, $x=10$ $\mu m$, $\tau_E=20$ ms.}
\label{radiometricparams1}
\end{figure}
Figure \ref{radiometricparams1} shows the single-pixel signal and background photoelectrons for a 1 cm debris as well as the cumulative (0-t)  $SNR^L$ (red line). We used $m_b=$20.0 mag arcsec$^-{2}$, a conservative value considering the observed range of 20-22 mag arcsec$^-{2}$ (European Southern Observatory - Paranal) \cite{patat-2003}. $SNR^L$ reaches an easily detectable level of 14 at the end of the track. The track passed through 20768 pixels and accumulated a total of 180 s of effective signal photon exposure. The single-pixel background photon exposure was fixed at $\tau_E=$20 ms. Therefore, the accumulated background photon exposure is ${\rm 20768 ~pixels} \times 20 ~{\rm ms ~per ~pixel}  = 415 s$. Similarly, the read noise is proportional to the number of pixels and is given by $e_n^2 = {\rm 20768} \times e_{n0}^2$. Here, we used $e_{n0} =1$ RMS.

 \begin{figure}[H]
  \centering
  \begin{minipage}[b]{0.4\textwidth}
    \scalebox{.9}{\includegraphics[width=8cm]{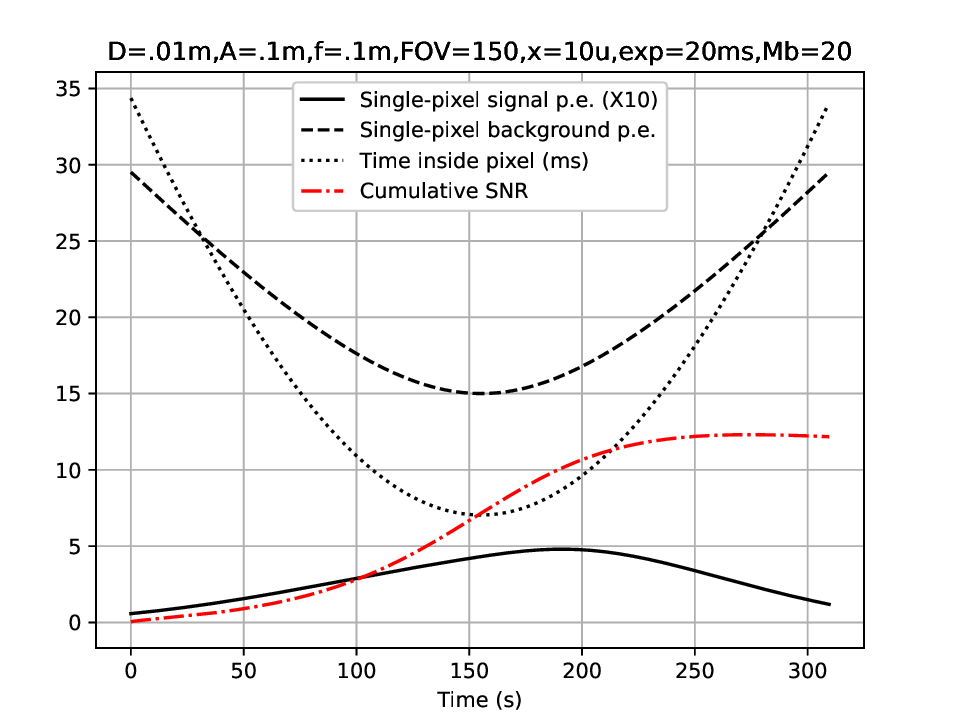}}
    %\caption{Caption Figure 4}
  \end{minipage}
  \hfill
  \begin{minipage}[b]{0.4\textwidth}
   \scalebox{.9}{\includegraphics[width=8cm]{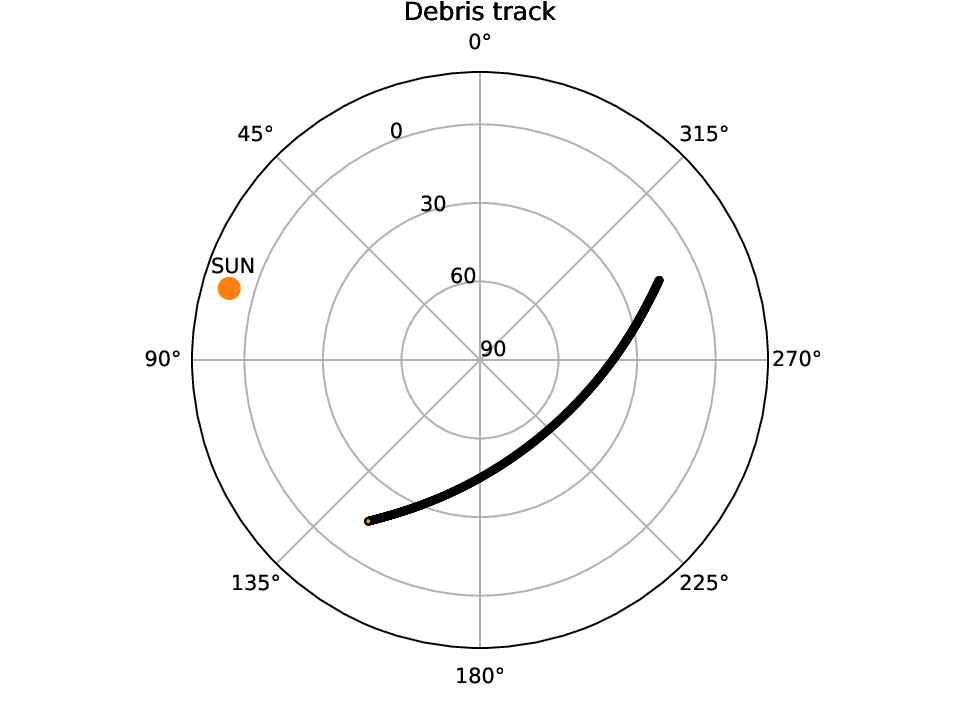}}
   % \caption{Caption Figure 5}
  \end{minipage}
  \caption{The same pass as Figure \ref{radiometricparams1} but viewed from a 4$^{\circ}$ higher location in latitude.}
  \label{4deghigher}
\end{figure}

% Clarification needed for figure 4 versus 5, ect

Figure \ref{4deghigher} (right) shows the same debris pass but viewed from a location 4$^{\circ}$ higher in latitude. This time, the section of the track with elevation $15^{\circ}+$ is shown. The duration is 620 s and the debris was sunlit throughout the track shown. Although being $\sim$3.5 times longer, $SNR^L$ reaches only $\sim$7 approximately half of
 $SNR^L$ for the overhead pass shown in Figure \ref{radiometricparams1}. Inspecting both figures, $R^2$ drop in the number of signal photons and higher background at lower elevations are responsible for the lower $SNR^L$ of the low-elevation pass. In addition, it can be seen that $SNR^L$ increases the fastest near the peak elevation. 

Note that in performing the integrations, it is assumed that the track is going through the center of pixels such that cross-track neighbors have no contribution. In reality, this is not the case because of the finite width of the point spread function (PSF), and possibly double or triple the number of pixels involved. In that case, the effective photon exposure of the integrated signal will remain the same, but the integrated background  will increase proportionally to the number of pixels involved, resulting in a reduction $SNR^L$ by a factor of 1.5-2. 

%Such reduction in $SNR^L$ may be avoided if the readings of the pixel boxes through which a track passes are weighted. For %example, for "square" pixels, if the track goes through the center of pixel, it should be assigned full weight. The other %extreme case, if the track is tangential or clips the corner of a pixel such that its net effect is a reduction of $SNR^L$, %that pixel should be discarded. Overall, the contribution of a pixel should be considered based on the proximity of their %centers to the track, the duration of the track segment over that pixel, the background emission at the location of the %pixel, read noise of the pixel, and the range of the track segment to the camera. 

Finally, for completeness, the $SNR^{L,M}$ of a cluster of $M$ identical cameras placed adjacently and looking in the same direction is given by the following. 
\begin{equation}
    SNR^{L,M} = \frac{M e_s^L}{\sqrt{M e_b^L+M(e_n^L)^2}} = \sqrt{M} SNR^L
\end{equation}
If one is to pursue increasing radiometric sensitivity to extreme limits by deploying such clusters in massive numbers, the detectable debris diameter will decrease with the fourth power of the number of cameras in the cluster.
$$
D_{detectable} \propto M^{1/4}
$$

 \begin{figure}[H]
  \centering
  \begin{minipage}[b]{0.45\textwidth}
    \scalebox{.9}{\includegraphics[width=7cm]{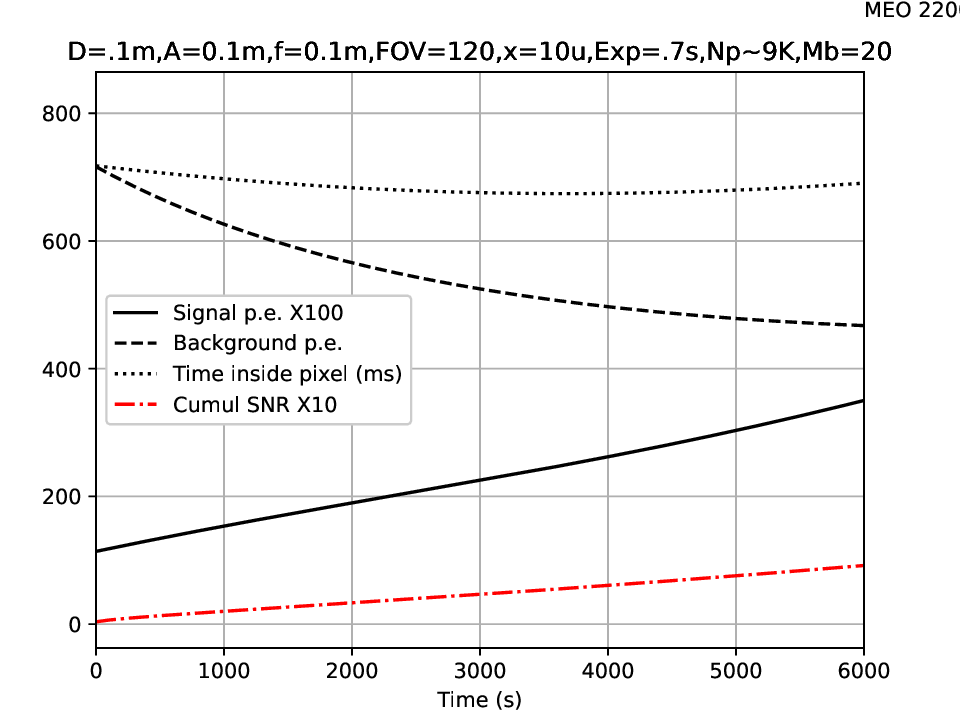}}
    %\caption{Caption Figure 4}
  \end{minipage}
  \begin{minipage}[b]{0.45\textwidth}
   \scalebox{.9}{\includegraphics[width=7cm]{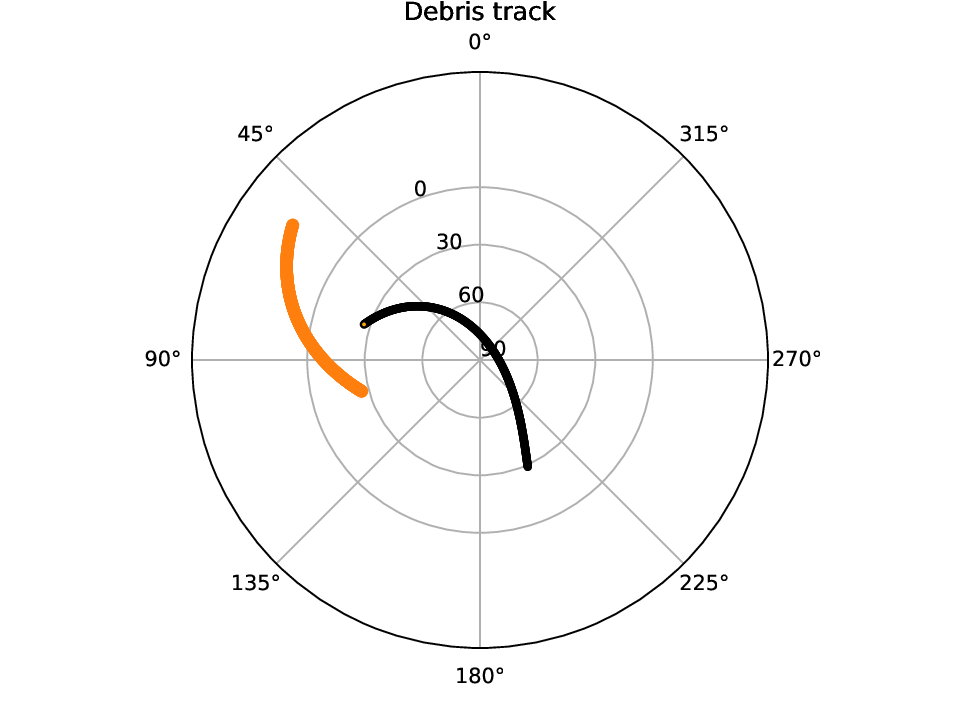}}
   % \caption{Caption Figure 5}
  \end{minipage}
  \caption{MEO orbit (22000 km). Cumulative $SNR^L$ (red dashed line) for a 10 cm debris 
 in MEO (22000 km). The trajectory corresponds to a GPS satellite TLE (Navstar 56).
 The orange line is the Sun's trajectory. Also shown are single-pixel SNR (solid black line), background (dashed black line) and time spent inside a pixel (dotted line). The optical parameters are: $d_{aperture}=10$ cm, $f=10$ cm, $x=10$ $\mu m$, $\tau_E=700$ ms.}
   \label{figure_meo}
\end{figure}
So far, we have considered LEO. Higher orbits are slower in angular rate, thus exposure times per pixel will be longer. Figure \ref{figure_meo} shows the cumulative SNR of a 10 cm m debris in  Medium Earth Orbit (MEO) with the same telescopic parameters used so far. The simulated orbit is that of the GPS satellite navstar 56. Note that the exposure time (500-700 ms) is significantly longer than that of LEO (10 ms). Although the SNR of a single-pixel is between $\sim 0.05-0.2$ (not shown in the plot), $SNR^L$ reaches $\sim 8$ after integrating about $\sim$9000 pixels over $\sim$1.6 hours.  While such hour-like long line integrals sound daunting computationally, they are simply slow-motion versions of the LEO integrals, with the number of pixels per integration remaining the same. However, the projection line forms are significantly different, and how it affects the computations will be investigated in a separate study.

 \begin{figure}[H]
  \centering
  \begin{minipage}[b]{0.45\textwidth}
    \scalebox{.9}{\includegraphics[width=7cm]{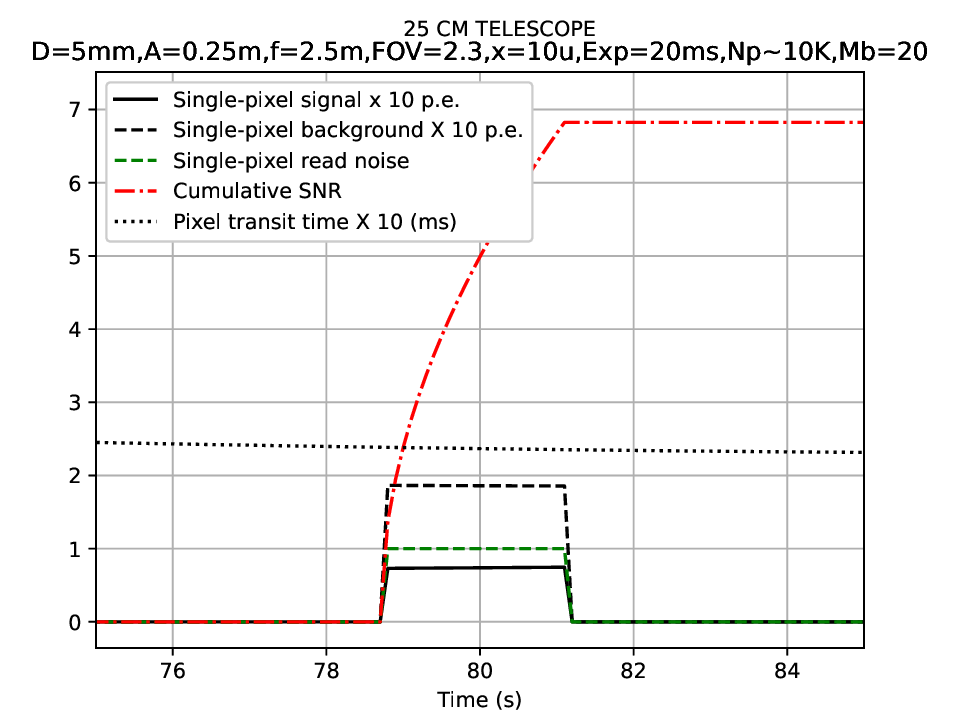}}
    %\caption{Caption Figure 4}
  \end{minipage}
  \begin{minipage}[b]{0.45\textwidth}
   \scalebox{.9}{\includegraphics[width=7cm]{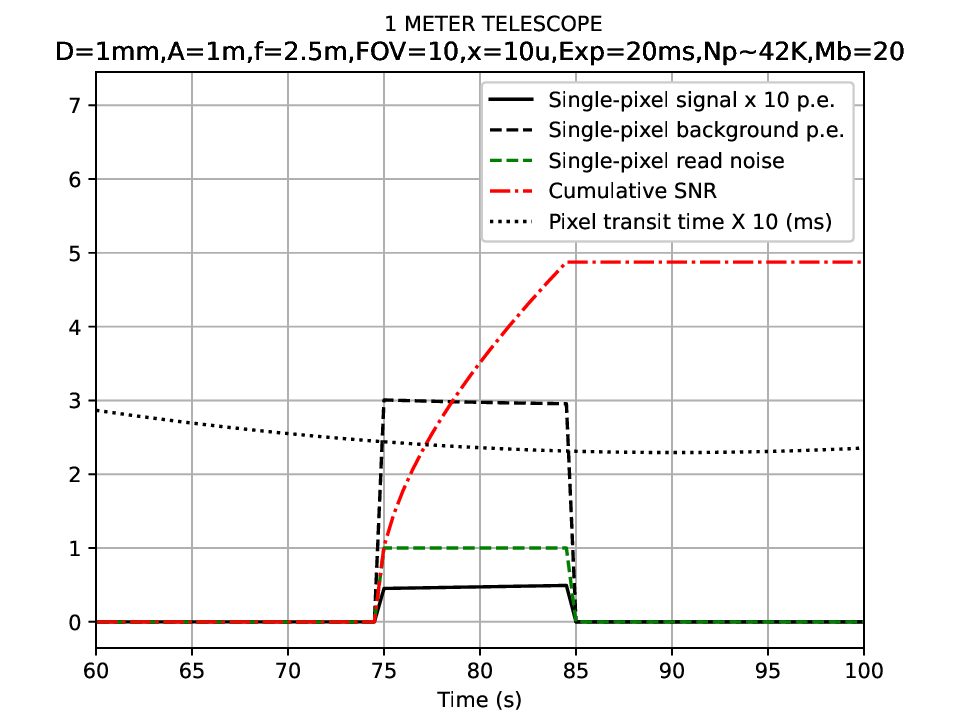}}
   % \caption{Caption Figure 5}
  \end{minipage}
  \caption{Cumulative SNR for an array of 25-cm (left) and an array of 1-m telescopes (right) for the orbit shown in Figure \ref{high_pass}. The number of telescopes are chosen to provide FOVs of 2.3$^{\circ}$ and 10$^{\circ}$, respectively. A constant exposure time of 20 ms is used for both systems.}
   \label{big_ones}
\end{figure}

We have shown thus far that it is theoretically possible to detect and track realistic size operational spacecraft ($> 10 {\rm cm}$) from LEO to GEO and objects as small as 1 cm in LEO with a modest 10 cm aperture telescope computing phase-space pixels.  Ultimately, we are interested in an SSA solution with a sensitivity to detect and track sub-cm debris, and that solution will require more aperture. 

Figure \ref{big_ones} shows the SNR performance for an array of 0.25m (left) and 1m (right) aperture telescopes that provide significantly greater sensitivity to detect 5 mm and 1 mm debris, respectively. The same orbit as in Figure \ref{high_pass} is used except the FOV is 2.3$^{\circ}$ for the 25 cm system and 10$^{\circ}$ for the 1-m system. Note that both systems have the same iFOV, i.e. $x/f$. Although the exposure times are fixed at 20 ms, the single-pixel background photoelectrons are different because of the difference in the aperture area $\times 16$. For the 25-cm system, the background is about 0.3, a small fraction of the read noise, whereas for the 1-m system it is 3X read noise.

 \begin{figure}[H]
  \centering
  \begin{minipage}[b]{0.45\textwidth}
    \scalebox{.9}{\includegraphics[width=7cm]{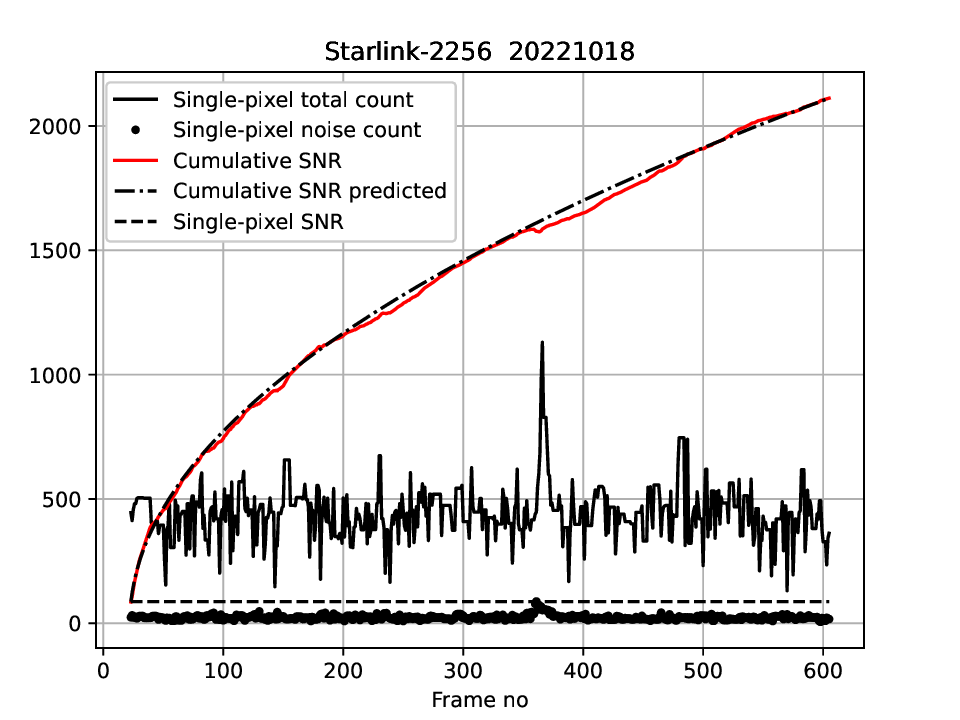}}
    %\caption{Caption Figure 4}
  \end{minipage}
  \begin{minipage}[b]{0.45\textwidth}
   \scalebox{.9}{\includegraphics[width=6.5cm]{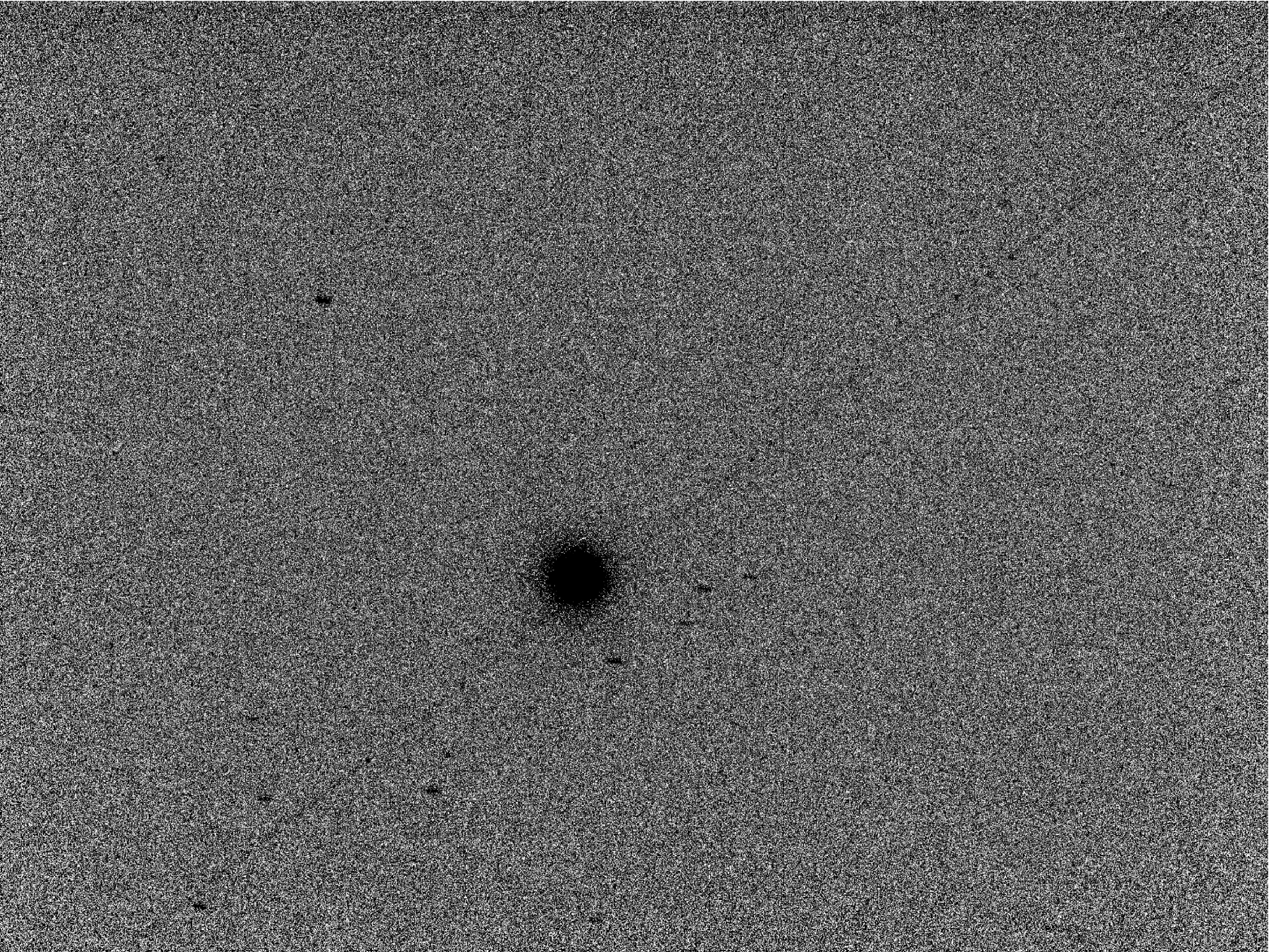}}
   % \caption{Caption Figure 5}
  \end{minipage}
  \caption{Starlink-2256 captured by a single William Optics telescope at 200 FPS. (Left) The cumulative SNR (red curve) matches well the predicted SNR (based on the $\sqrt{N}$ rule). Shown also is the single-pixel SNR obtained from averaged single-pixel signal and noise values. (Right) Average of $\approx$600 frames corresponding to an effective exposure of 3s. The faint diagonal line is the satellite passing by Sirius, the brightest star in the night sky.}
   \label{starlink}
\end{figure}

Figure \ref{starlink} shows data for a Starlink-2256 pass captured by a single 250mm focal length, 51mm diameter aperture William Optics telescope (1.2$^{\circ}$ FOV ) on 10/18/2022.  The images were captured at a framerate of 200 fps and the object was located ahead of time using the planetarium software, Stellarium. A single-pixel-detection based tracking algorithm was developed to find the $(i,j)$ path. The signal was integrated along this path. The cumulative SNR increases proportional to $\sqrt{N}$ as predicted. The image on the right is an average of all the frames throughout the 3s pass and the Starlink track is distinguishable as the diagonal line.

\section{Phase-space-pixel resolution and the search space}
We defined a phase-space-pixel or PSP as the sum of the values of all the pixels that line up on the projection of a hypothetical phase-space trajectory on the camera FOV. However, the terminology {\it phase space} can be used for any multidimensional space; it does not have to be a space of  position and velocity only. Similarly, as $[i,j]$ uniquely identifies a pixel, here we see an orbit as a unique point in a six-dimensional phase space. If we can properly define the resolution of that one point, we can calculate the orbital uncertainty and the number of grains that we have to sort through to find the object.

The resolution of PSP can be understood as follows. In two-dimensions, the angular resolution of a conventional pixel camera can be defined by the angular extent by which two closely-spaced point sources in $\mathbb{R}^{2}$ can be distinguished.  It corresponds to a grain of steradian at an azimuth and elevation point. In four-dimensions, the resolution of a vector-pixel, as used by \cite{heinze+etal-2015}, can be defined as the extent to which two closely-spaced constant-angular speed point sources in angular phase-space can be distinguished. This is how MEO-GEO or Near-Earth-Orbit (EO) observations are modeled. The object has two angular position and 2 angular speed variables and one performs a search in the $\mathbb{R}^{2}$ space by computing vector-pixels as the sum of pixels corresponding to an angular phase-space trajectory.  For long integration times over a wide FOV, a vector-pixel does not have enough dimensions to describe the track as the angular speed changes within the FOV.   We therefore compute PSP over a six dimensional phase-space trajectory parametrized by the orbital elements. Then, the PSP resolution  can be defined as the extent to which two closely-spaced accelerating point sources in $\mathbb{R}^{6}$ can be distinguished.

The PSP resolution can be computed as follows.  A point source with a hypothetical orbit excites photoelectrons on a line of pixels, which can be termed a "line spread function" or LSF such that the cross-track width is given by the point-spread function (PSF). For simplicity, we will assume PSF is a delta function; therefore, LSF is an infinitely thin stripe. This means that only one pixel, the one whose boundaries contain the debris, is activated at a time. The PSP resolution can therefore be defined as the 6D grain size at a given position in the $\mathbb{R}^{6}$ space.  The grain size can also be interpreted as the  orbital uncertainty defined as the coarseness of the orbital parameters that are consistent with the given measurements \cite{folcik+etal-2011}. 

The boundaries of the grain can be defined so that, throughout the volume of a grain, half of the projection pixels remain the same. To determine the 6D grain size for an orbit ${\bf x} = [\theta, \Omega, e, \omega, M, r]^T$ that starts at time $t_0$ and ends at $t_1$, we first compute the corresponding pixel positions
\begin{equation}
(i,j,k) \in L_{t_0}^{t_1}({\bf x})
\label{pixelpositions}
\end{equation}
where $[i,j]$ are the row and column positions of the $k$th frame and $L$ is the LSF. The $\mathbb{R}^{6}$ parameters above are inclination ($\theta$), longitude of the ascending node ($\Omega$), eccentricity ($e$), argument of the perigee ($\omega$), mean anomaly ($M$) and revolutions per day ($r$). If the time interval $t_0-t_1$ is such that the orbital segment is outside the FOV, $(i,j,k)$ will be an empty set ($\varnothing$). Generally, we can choose $t_0, t_1$ to correspond to the entry and exit times on the FOV perimeter pixels. Moreover, to maximize exposure time $\delta_\tau=t_1-t_0$, we can avoid computing trajectories that pass marginally through the FOV, such as those that clip the corner of the imaged area. For the following analysis, we assume a trajectory that is fully contained in the FOV. To compute the resolution in $\theta$, we first compute the set 
$$A=(i,j,k) \in L(\theta, \Omega, e, \omega, M, r)$$
Then, we compute the set for the same trajectory but increment by $\Delta\theta$ in inclination:
$$B=(i,j,k) \in L(\theta+\Delta\theta, \Omega, e, \omega, M, r)$$ The resolution in inclination is given by the increment that results in exactly half the pixels common to both trajectories, i.e.,
$$\theta_{resolution} = arg_{\Delta\theta} n(A \cap B) = n(A)/2.$$
We do the same procedure to compute the resolution for the other dimensions. 

Here we compute resolutions for a test orbit for the camera system discussed above. We used an NxN pixel camera pointed at zenith and located the observatory at the equator at 0$^{\circ}$ longitude. We will use a test orbit that overflies the observatory at $h=420$ km at a speed of 7672 m/s and passes through its zenith at $t=0$. The test orbit has the following parameters:
$\theta=90, \Omega=360.0, e=0, \omega=0, M=0, r=15.49962$, which corresponds to a circular orbit of 90$^{\circ}$ inclination at an altitude of 420 km. Due to the rotation of Earth, the test object approaches the zenith from the azimuth of 176$^{\circ}$ and departs from the azimuth of 356$^{\circ}$. 

We now change each of the 6 orbit parameters in small increments until we find the increment that corresponds to a set $B=(i,j,k)$ such that only 50\% of the pixels are common with the original set $A=(i,j,k)$ We do this for $N=10, 100, 1000, 10000$. Table \ref{table_grain}  shows the grain sizes for all orbital parameters with the exception of $e$ and $\omega$, since we assumed the test orbit to be circular.  We can then arrive at the following analysis. If the track can be approximated as a straight line in the image, the orbit inclination ($\theta$) corresponds to the slope of the track, and longer tracks (higher $N$) give higher resolution.  Since the slope on an $NxN$ frame  can be determined roughly with a precision of $1$ pixel over $N$ pixels, the numbers for $d\theta$ for different $N$ are roughly given by $\arctan(1/N)$. Next, an increment of $\Omega$ corresponds to a shift of the track to the left or right, and, again assuming a straight track,  the resolution d$\Omega$ is independent of the track length and roughly corresponds to a shift $\Omega$ that corresponds to lateral movement of the track by half a pixel. Therefore, the spatial resolution of the cross-track can be determined at a resolution of $(x/2f) \times h$ = 21 m. Next, the resolution of $r$ is inversely proportional to $N$. This is so because the angular rate, which is proportional to $r$, can be determined with a precision of $\tau_e/N$, where $\tau_e$ is the exposure time or the inverse of the frame rate. The values of $dr$ in Table \ref{table_grain} correspond to 23, 8.7, 0.87, and 0.087 km, respectively. Finally, the mean resolution of the anomaly is determined by $\tau_e$. Here we used an exposure time of $5$ ms. Considering the 92.9 min orbit here, we get $360.0^\circ \times 5 ms /(92.9 min \times 60 s)  = 0.000161^\circ$, which closely matches $dM$ in Table \ref{table_grain}. This means that the track position can be determined with a resolution of  7672 m/s $\times$ 5 ms = 38 m.

\begin{table}
\centering
\begin{tabular}{|c c c c c|} 
 \hline
 N (pixels) & 10 & 100 & 1000 & 10000\\ [0.5ex] 
 \hline\hline
 $d\theta$ (deg)& 3.04 & 1.21 & 0.13 & 0.013\\ 
 \hline
 $d\Omega$ (deg) & 0.000180 & 0.000176 & 0.000175 & 0.000174\\
  \hline
 $dr$ (rev/day)& 0.079 &  0.030 & 0.003 & 0.0003\\ 
 \hline
 $dM$ (deg) & 0.000175 &  0.000175 & 0.000175 & 0.000187\\ 
% \hline
% $N_T (tracks/s)$ & 2.6e6 & 1.7e7 & 1.6e10 & 1.6e13\\ 
 [1ex] 
 \hline
\end{tabular}
\caption{\label{table_grain} Computationally-obtained track resolutions for a northbound LEO test circular orbit in LEO overflying through the zenith of an optical observatory located at the equator for a range of $NXN$ camera systems with the same IFOV. $dr$ can be converted to altitude resolution $dh$ by multiplying it by $\approx 300$, e.g., $dh\approx 0.1$ km for $N=10000$. 
%The bottom row shows the number of trajectories to be computed by the imager within the ranges of 90$^\circ$ %inclination and 100 km altitude range centered at 420 km for each second of observations.
}
\end{table}

% \hline
% $de$ & 0.0010052 & 0.0010485 & 0.0001276 & 0.0000146\\
% \hline
% $\omega$ & NA & NA & NA & NA\\

%which is comparable to the number of circular orbits for the full search %case above with N=10000 but significantly smaller if all orbits considered.

%Dividing the 1 min satellite orbit into 1 km hazard segments, we obtain 8 km/s*60s/1 km = 480 %intercept points. Going back to the algorithm above,

%Moreover, assume one particle is launched from every perimeter pixel every 10 ms. Number of %trajectories to be calculated every second: 40000 perimeter pixels  x 1000 intercept points x 100 %frames= 4e9 trajectories. 

%4. If we run the code to protect all 10000 operating satellites, we need to compute 4e13 %trajectories every second. Compare this to 1e17 trajectories per second for full search. That is %interceptor search accounts for .04 percent of full search. 

\begin{figure}[H]
\centering\includegraphics[width=10cm]{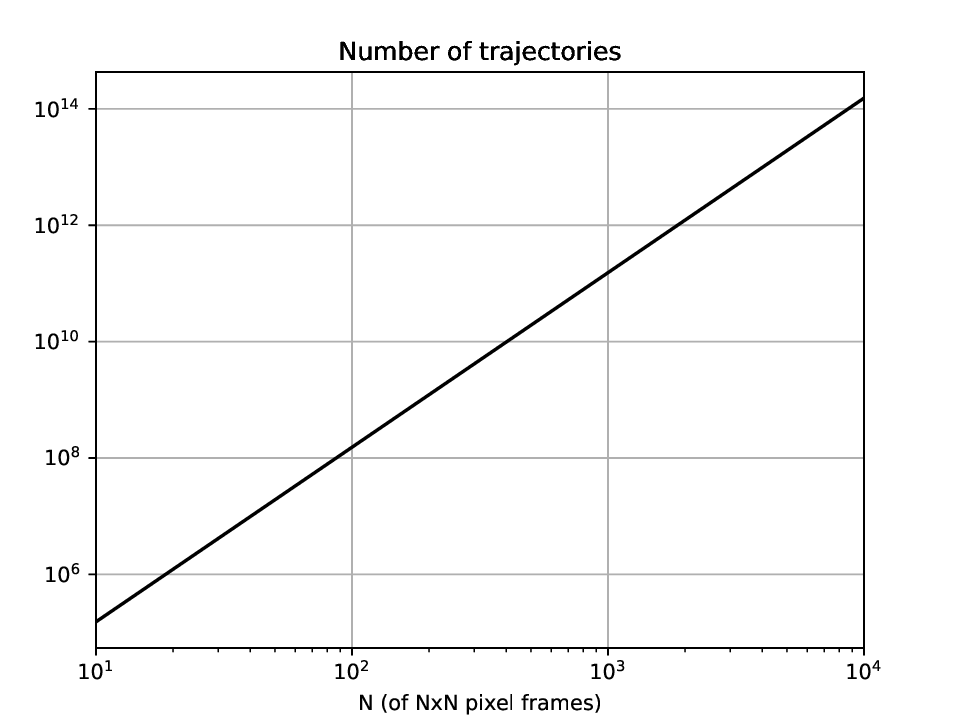}
\caption{Number of all possible distinct straight tracks that can exist at any given moment. Tracks start at a perimeter pixel of an $NXN$ frame and exit at a perimeter pixel of another frame at an angular rate with a lower and upper bound.  The exposure time is set to IFOV divided by the angular rate for a circular orbit at 400 km altitude so that a debris at 400 km transits exactly one pixel per frame). The angular rate is restricted to circular orbits between the altitudes of 350-450 km. Lower (higher) altitude debris transits faster (slower). For example, a 400-km debris entering at the corner of an 100x100 frame and transiting diagonally will exit the FOV at the opposite corner of the 141st frame, while a 450-km debris starting on the same path will exit at the 158th frame.}
 \label{ntrajectories}
\end{figure}
We now calculate the search space. A simple estimate is to constrain the search to trajectories short enough to be considered straight. This means that the angular rate along a trajectory is constant and the phase-space pixel is 4-dimensional. A debris entering the FOV at a given frame's perimeter pixel and at a given altitude and direction can exit the FOV later only at a perimeter of a particular frame. We therefore count only the entry-exit pixel pairs whose corresponding angular rate falls within the limits of the circular orbit region being searched. Figure \ref{ntrajectories} shows the number of such tracks in the altitude range 350-450 km (circular orbit). At a given time, some tracks are initiated at the perimeter, while some tracks reach the perimeter and are terminated. The numbers in Figure \ref{ntrajectories} are those of equilibrium.

However, the search space could be significantly smaller if we look only for debris intercepting an operating spacecraft whose orbits are well known. We can use SGP4 to find hypothetical debris TLE’s that begin within the FOV of the observatory and intercept a particular operational satellite at a future time interval.  No assumption of orbital circularity is needed. 

Consider that the time now is $t_0$ and an operational satellite will be in position ${\bf r}_{sat}$ at time $t_0+\Delta t$. Consider also that at $t_0$ a debris enters the FOV at the perimeter pixel ($i,j$) in the frame $k_0$, corresponding to the position (${\bf r}_{deb}$). Since we do not know the range to the debris, it will be one of the search variables.  We now invoke Lambert's problem: determine the interceptor debris orbit ${\bf x}_i$ from the two position vectors (${\bf r}_{sat}(t_0+\Delta t)$, ${\bf r}_{deb}(t_0)$) and the time of flight $\Delta t$: 
\begin{equation}
{\bf x}_i=[\theta,\Omega,e,\omega,M,r]^T = W({\bf r}_{deb},{\bf r}_{sat},\Delta t)
\end{equation}
where $W$ is the Lambert's function.  Then, we apply the following example algorithm for LEO at every frame:
\begin{verbatim}
for dt in range(min_warning_time, time_inc, max_warning_time):
    for each perimeter (i,j) and debris_range in (r_min, dh, r_max):
        r_deb = debris_position(i,j,debris_range)
        r_sat = OrbitPropagator(TLEsat, t0+dt)
        x_i = LambertsFunction(r_deb, r_sat, dt)
        if x_i belongs to LEO:
            [i,j,k] = compute_track_over_fov(x_i)
            SNR = integrate_over([i,j,k])
            if (SNR > SNR_threshold)  
                issue warning
                catalog x_i
\end{verbatim}
Here, the satellite path is divided into segments sufficiently small for the intercept to be considered a hazard. However, only a small fraction of an orbital period is interceptable. For the rest, the debris simply cannot "catch up", unless its orbit is very highly eccentric or parabolic, which could place it out of the LEO category. For example, neglecting Earth's rotation, a debris departing northbound overhead our optical observatory located on the equator at 0$^{\circ}$ longitude will intercept a similar altitude northbound satellite ascending at 90$^{\circ}$ longitude  at the north pole in 22.5 minutes, assuming 90 min orbital period.  If we restrict the debris altitude to within $\pm$50 km of the satellite altitude, its speed must be within 0.37 percent of the satellite's speed. Within this speed range, the particle can intercept the satellite as early as 5 s (from $22.5 min \times \Delta v/v$) before it passes the north pole or as late as 5 s after, an interval of 10 s, which is a small fraction ($\approx.2\%$) of the orbital period (90 min).

Now we calculate the number of distinct trajectories that intercept an operational LEO satellite. For each of the perimeter pixels (a total of $4N$) and for each resolvable altitude bin $dh(N)$ (see Table \ref{table_grain}) in an altitude range $H$,  we assume that an intercept exists only for $T$ seconds per orbital period. If $\Delta x$ is the distance below which it is considered an intercept, the number of intercept segments per orbit is $v_{sat}T/ \Delta x$ where $v_{sat}$ is the orbital speed. However, we must consider the system resolution. Considering the north pole intercept scenario above, the imager can only resolve limited segments of the target satellite. The length of a resolvable segment can be approximated as $\Delta x = v_{sat} dt d\theta (N)$. From Table \ref{table_grain},  for $N=1000$, $d\theta=0.13^{\circ}$. Propagating the orbit for $dt=22.5 min$, we obtain $\Delta x=23$ km. $\Delta x$ will expand further if the intercept is searched for longer warning times. The number of intercept segments is then $(v_{sat}T) / (v_{sat} dt d\theta (N)) = T/(dt d\theta(N))$
Now, the number of intercept-only trajectories $N_{IT}$ per second can be expressed as:
$$
N_{IT} = 4N  \times (1/\tau_e) \times (H/dh(N)) \times (T/(dtd\theta(N))
$$
Note that the factor $1/\tau_e$ is frames per second and there are projectiles launched from the perimeter pixels in every frame.  Assuming the following parameters, $N=1000$, $\tau_e=0.01$s, $H=100$km, $dh=1$ km, $T=10$s, $d\theta=0.13^{\circ}$ and $dt=22.5$ min, we obtain $N_{IT}=1.3e8$/s. However, this number is used to protect only one satellite. It must be multiplied by the number of satellites to protect a fleet.

Another possibility of narrowing the search space is to search for glints. This mode of detection is based on the hypothesis that at various points along the debris trajectory, there will be a burst of specular reflections off the debris with the number of photons per pixels reaching orders of magnitude larger  than the average per pixel. Such glints would produce short but bright tracklets that can be detected with the aforementioned straight-line integrations but small $N$. Moreover, because of the likely rotation of the debris and the changing reflection angle, the glints may repeat, producing a set of tracklets within the FOV. As discussed in the following on orbit determination, once one tracklet is detected, it can be linked with the rest in the FOV as part of the initial orbit determination. It is also possible to link tracklets from multiple passes using similar tracklet linkage algorithms developed to observe small NEOs with the Space Surveillance Telescope \cite{lue+etal-2019} and in many others. However, the detection methodology used by Lue et al. [2019] differs from ours, as the tracklets were single frame detections whereas ours are track-before-detects, allowing us to have enhanced sensitivity.

Note that MEO and GEO will differ from LEO mainly by the time scales and projection trajectory shapes. The debris will spend significantly more time within the FOV. The exposure time will be 100x to 1000x longer.  The frame rate and the data throughput will be less by the same factors. However, the trajectory shapes, far from being straight, will be curved lines or spirals. The exposure rates need to be optimized for these slow motion regions, however. Short exposures used for LEO may amount to unnecessarily high read noise for MEO and GEO. Nevertheless, if some loss of performance can be tolerated, the same collected data can be mined for debris everywhere from LEO to GEO.

\section{Orbit determination}

In this section, we briefly discuss an analytical framework for Initial orbit determination (IOD) using the SPOT analysis. IOD  generally refers to the direct computation of six orbital elements from detections during routine data mining for uncatalogued debris. Most IOD methods are based on solving six equations for six orbital elements from two position measurements or three sets of angle measurements. \cite{montenbruck+gill-2000} In SPOT, however, image processing for measurements and orbit determination is inherently coupled and is therefore treated as a single complete problem. In particular, because the measurements are integrals, a measurement is not complete until the trajectory in the FOV is estimated, and the trajectory cannot be fully estimated until the measurement is complete. The brute-force search approach of integrating over every possible full trajectory is not computationally feasible. Therefore, in the following, we take a {\it recursive measure and fit} method with track extension at every iteration.  Once all iterations are complete, the result corresponds to a single measurement corresponding to a single pass through the FOV. Another note, for the purpose of modeling the FOV projection below, we will describe an orbit with six-parameters only, neglecting non-centric gravitational and non-gravitational forces that must be included for orbit propagation. However, for a passage time of several minutes over the FOV, the corresponding accelerations for cm and larger objects are too small to cause a deviation in the set of pixels. For sub-cm debris, however, these effects must be properly accounted for, both inside the FOV and outside it for orbit propagation.   

First, a phase-space pixel (PSP) is a vector containing a set of parameters from which we can determine, directly or through a coordinate transform, a trajectory that passes through the pixels of a datacube. The PSP intensity value is the sum or weighted sum of these pixel values. A detection is a PSP intensity that exceeds a certain signal-to-noise ratio (SNR) threshold with a corresponding probability of occupancy (POC). There is only one detection event per debris per pass, although neighboring PSPs may light up, just like multiple pixels light up around a point source due to the point spread function (PSF) of the optical system.

A PSP can correspond to short straight lines, intermediate-length low-order polynomials, or long curved lines as direct space-time FOV projections. The track length depends on how sensitive we want the search to be, as the exposure time increases with the track length. The most sensitive search involves whole FOV integrations over exact TLEs projections, which directly yields TLEs. However, this is also computationally the most expensive. On the other hand, when short segments return a detection, the associated orbital uncertainty could be large as seen in Table \ref{table_grain}. Nevertheless, once a short segment is detected, it can be "pulled like a thread" to find the rest of it, with the SNR growing as it is pulled, eventually the length spanning the entire FOV, resulting in a precision TLE with a high SNR.

Let ${\bf z}$ denote a straight-line measurement, which is the PSP position of a detection event, and assume that there are no measurement errors. ${\bf z}$ is equivalent to the measured angular position, e.g. $ { \bf \theta} =[\theta_{az}, \theta_{el}]^T$  of a point source on a conventional image or to the measured radar range $r$ of a point target. Let us say that the search using short straight lines returned a detection at ${\bf z}_0=[i_0, j_0,k_0,\alpha_i, \alpha_j, \alpha_k, t_1, t_2]^T$ whose corresponding track pixel positions $(i,j)$ before rounding are given by $[i=i_0 + \alpha_i t, j=j_0 + \alpha_j t, k=k_0 + \alpha_k t]$, where $k$ is the frame number, $(\alpha_i, \alpha_j, \alpha_k)$ are slopes of ($i,j,k$) progressions and $(t_1, t_2)$ are the track start and end times. For example, a stationary source would be represented by ($\alpha_i=0, \alpha_j=0, \alpha_k=1/\tau_e$), while streaks on single frames would be represented by $\alpha_{i,j} \gg \alpha_k$. 

We now find in a library $\mathcal{L}$ of coarse-grained orbits a guess orbit ${\bf x}_0=[\theta_0,\Omega_0,e_0,\omega_0, M_0, r_0]^T$ that corresponds to the detected straight line, i.e.,
$$
{\bf z_0}= h[{\bf x}_0],~~~~ {\bf x} \in \mathcal{L}
$$
where $h$ is a transform that projects ${\bf x}$ to ${\bf z}$ in space-time. 
The grain sizes in $\mathcal{L}$ are such that there is one-to-one correspondence between every ${\bf z}$ and ${\bf x}$. For very short tracks, the six-dimensionality of ${\bf x}$ may be reduced to four by setting ($e=0, \omega=0$) corresponding to circular orbits.

The next step is to "pull" the newly found thread by repeating the search with iteratively longer tracks beginning with ${\bf x}_0$. This means we are bringing in more measurements into the estimate, until we reach the end of the track when the object is out of the FOV.
For this, we define the PSP intensity function of track length $L_m$ for iteration $m$,
$$
Q^{L_m}(h({\bf x})) = \sum_{(i,j,k) \in h({\bf x})}^{L_m} w_{i,j,k}I_{i,j,k} 
$$
which is a weighted sum of the values of the pixels ($i,j,k$) in the projection of ${\bf x}$ in the FOV. The coupled orbit estimation/signal intensification process now attempts to deduce a highly resolved value of ${\bf x}$ that maximizes $Q$. A necessary condition for the intensity function to be maximum with respect to ${\bf x}$ is that 
$$
\frac{ \partial Q^{L_m} } { \partial {\bf x}}=0
$$
A simple solution is to apply the gradient ascent method. From ${\bf x}_0$,  we take steps proportional to the positive gradient of the function at the current point:
$$
{\bf x}_{n+1} = {\bf x}_n +  \lambda \nabla Q({\bf x}_n)
$$
The gradient descent will quickly approach the solution from a distance, but
its convergence will slow down when it is close to the solution. At some point,  the track length will increase to $L_{m+1}$ and we repeat the algorithm starting from the last estimate of ${\bf x}$.
This effect of increasing the length of the track can be visualized as gradually steepening the $Q$ landscape to prevent the algorithm from "meandering". 
Once $L$ is long enough to span the entire FOV, the final fit ${\bf x}_n^{final}$ becomes the solution. In other words, after using all the relevant data for a single pass, our system reports a detection at a particular PSP with the measurement value ${\bf x}_n^{final}$. 

We now calculate the Jacobian for the covariance matrix to determine the uncertainty of the estimated orbit. The above discussion assumes the absence of errors. With errors we have
$$
{\bf x_n^{final}}={\bf x}_{true}+{{\bf \epsilon}(L)}
$$
where ${{\bf \epsilon}(L)}$ represents measurement errors that depend on track-length, as well as on several other factors, including SNR, atmospheric diffraction, or optical aberrations.   Just as we can assign a conventional pixel an angular accuracy or a radar range measurement a range accuracy, we can do the same for a PSP by assigning each of its parameters an uncertainty or we can work with ${\bf \epsilon \epsilon}^T$ to account for correlations between the measurement components. The statistics of measurement errors can be obtained by taking multiple PSP measurements of highly visible spacecraft with precisely-known orbits, e.g. via GPS, then comparing those measurements to the more-precise orbital parameters.  If we assume that ${\bf \epsilon^L}$ is zero-mean Gaussian, we  can form the weighting matrix as ${\bf W}=(E(\epsilon \epsilon^T))^{-1}$. The orbital uncertainty associated with the final solution ${\bf x}_n^{final}$
can now be computed from the covariance matrix 
$$
C=({\bf J^TWJ})^{-1}
$$
where the Jacobian
$$
J=\frac{\partial Q}{\partial {\bf x} } |_{{\bf x}={\bf x}_n^{final}}
$$
contains the partial derivatives of the observations modeled with respect to the state vector at the reference epoch ${\bf x}_n^{final}$.
The diagonal elements of the covariance matrix yield the standard deviations of the elements of ${\bf x}$ and are a measure of the intervals that most likely contain the actual state.

\section{Conclusion}
The performance calculations using a geodesic line integral formulation show that sub-cm debris can be detected and tracked by a generalized Hough transform of data from arrays of cameras positioned to provide a wide FOV to maximize the light collected off debris during twilight hours. Compared to a single-pixel signal-to-noise ratio, the line integral significantly increases the SNR roughly proportional to the square root of the number of pixels along the track, implying the importance of synthetic exposure time. We show that such line integrals not only increase the sensitivity, but also provide meter-scale resolution in locating the object. However, a significant gain in sensitivity and resolution comes with a heavy computational cost, which requires the computation of trillions or more tracks per second. The computational task and orbit determination will be discussed in companion papers. 
%We conclude that a full-track-detect method, large number of camera arrays operating during clear twilight conditions, a %network of such observatories and GPU-based many-core parallel processing platforms can provide the performance needed to %catalog sub-cm diameter debris. 

\section{Acknowledgements}

This work was partially supported by the U.S. National Science Foundation grant no. 2404362 to OpticalX, LLC.

%\begin{backmatter}
%\bmsection{Funding}
%\bmsection{Acknowledgments}
%\bmsection{Disclosures}
%\noindent The authors declare no conflicts of interest.
%\bmsection{Data Availability Statement}
%A Data Availability Statement (DAS).
%\bmsection{Supplemental document}
%See Supplement 1 for supporting content. 
%\end{backmatter}

%%%%%%%%%%%%%%%%%%%%%%% References %%%%%%%%%%%%%%%%%%%%%%%%%

%%%%%%%%%% If using BibTeX:
\bibliography{radiometric}

%%%%%%%%%% If preparing manually:
% \begin{thebibliography}{1}
% \newcommand{\enquote}[1]{``#1''}

% \bibitem{Zhang:14}
% Y.~Zhang, S.~Qiao, L.~Sun, Q.~W. Shi, W.~Huang, L.~Li, and Z.~Yang,
%   \enquote{Photoinduced active terahertz metamaterials with nanostructured
%   vanadium dioxide film deposited by sol-gel method,}
%   {\protect\JournalTitle{Optics Express}} \textbf{22}, 11070--11078 (2014).

% \bibitem{Optica}
% {Optica}, \enquote{{Optica Publishing Group},}
%   \url{http://www.opg.optica.org}.

% \bibitem{FORSTER2007}
% P.~Forster, V.~Ramaswamy, P.~Artaxo, T.~Bernsten, R.~Betts, D.~Fahey,
%   J.~Haywood, J.~Lean, D.~Lowe, G.~Myhre, J.~Nganga, R.~Prinn, G.~Raga,
%   M.~Schulz, and R.~V. Dorland, \enquote{Changes in atmospheric consituents and
%   in radiative forcing,} in \enquote{Climate Change 2007: The Physical Science
%   Basis. Contribution of Working Group 1 to the Fourth assesment report of
%   Intergovernmental Panel on Climate Change,}  S.~Solomon, D.~Qin, M.~Manning,
%   Z.~Chen, M.~Marquis, K.~B. Averyt, M.~Tignor, and H.~L. Miler, eds.
%   (Cambridge University Press, 2007).

% \end{thebibliography}

\end{document}